# An Adiabatic Superconductor Comparator with 46 nA Sensitivity

Naoki Takeuchi, *Member, IEEE*, Taiki Yamae, Hideo Suzuki, and Nobuyuki Yoshikawa, *Senior Member, IEEE*

*Abstract*—Adiabatic quantum-flux-parametron (AQFP) circuits can operate with extremely small energy dissipation (~1 zJ per junction at 5 GHz) and high sensitivity (~1 µA) owing to adiabatic switching. Thus, AQFP logic is suitable to use as readout interfaces for cryogenic detectors, such as superconducting nanowire single-photon detectors (SSPDs). In order to extend the application of AQFP logic to various detectors, it is crucial to achieve even better sensitivity because some detectors, such as WSi SSPDs and transition edge sensors (TESs), require sub-µA sensitivity. In the present study, we propose a high-sensitivity AQFP comparator with a current transformer (CT), which increases the equivalent input current and thereby improves the sensitivity. Numerical simulation shows that the proposed comparator achieves a sensitivity approximately six times better than that of a conventional AQFP comparator. Furthermore, we demonstrate a sensitivity of 46 nA at 4.2 K for a sampling frequency of 500 Hz in the experiment.

## I. INTRODUCTION

SUPERCONDUCTOR digital circuits [1] can operate with low power dissipation at cryogenic temperatures and thus have been used as readout interfaces for cryogenic detectors [2–7]. Specifically, we proposed the use of adiabatic quantum-flux-parametron (AQFP) logic [8] as readout interfaces [9] for superconducting nanowire single-photon detectors (SSPDs) [10, 11] because AQFP logic is not only energy efficient [12], but also possesses physical features suitable for detector applications, as follows. AQFP gates are serially biased by ac excitation currents [13], so that the supply current (a few milliamps) does not increase with the circuit scale. This is important for cryocooler implementation because the supply current is limited to a few hundred milliamps by the cooling power for some compact cryocoolers [14], beyond which the supply current and parasitic resistance inside the cryocooler generate serious Joule heating and increase the sample stage temperature. In addition, AQFP gates have high sensitivity owing to adiabatic switching [15, 16], where the logic state can be switched by an input current that tilts the potential energy slightly. We demonstrated an AQFP comparator with a sensitivity of approximately 1 µA [17], which is small enough to digitize the signal current from an NbTiN SSPD (~10 µA). In fact, we have successfully demonstrated NbTiN SSPDs using AQFP circuits [18, 19], which digitize and encode the signal currents from the SSPDs in a cryocooler in order to reduce the number of coaxial cables required for reading out the SSPDs.

In order to extend the application of AQFP circuits to other cryogenic detectors, it is crucial to improve the sensitivity of AQFP circuits because some detectors, such as WSi SSPDs [20, 21] and transition edge sensors (TESs) [22, 23], require sub-µA sensitivity. In the present study, we propose a high-sensitivity AQFP comparator that uses a current transformer (CT) [4] to increase the equivalent input current and thereby improve the sensitivity. Numerical simulation verifies that the proposed comparator achieves a sub-µA sensitivity owing to the CT. Moreover, we experimentally demonstrate that the proposed comparator achieves a sensitivity of 46 nA at 4.2 K for a sampling frequency of 500 Hz.

## II. AQFP COMPARATORS

### A. Conventional Comparator

Figure 1(a) shows a circuit diagram of a conventional AQFP comparator, which works as a buffer in digital circuits. Assuming that an input current $I_{in}$ is applied by a detector such as an SSPD to the comparator via a 50 Ω interconnection [18, 19], $I_{in}$ is terminated by a resistor $R_t$ (= 50 Ω). It is also assumed that the impedance of the detector is much higher than $R_t$, so that $R_t$ does not change the value of $I_{in}$. A dc offset current $I_d$ applies a dc magnetic flux of $0.5\Phi_0$ to a dc superconducting quantum interference device (SQUID) composed of paired inductors ($L_1$ and $L_2$) and Josephson junctions ($J_1$ and $J_2$), where $\Phi_0$ is the flux quantum. An ac excitation current $I_x$ applies an ac magnetic flux with an amplitude of $0.5\Phi_0$ to the dc SQUID. The reason why $I_d$ and $I_x$ apply dc and ac fluxes separately via $L_d$ and $L_x$ is to operate the comparator by four-phase clocking [13], as will be shown in Sec. III. In general, the dc SQUID is symmetrical, so

The present study was supported by JSPS KAKENHI (Grants No. 18H05245, No. 18H01493, and No. 19H05614). *(Corresponding author: Naoki Takeuchi)*

N. Takeuchi and H. Suzuki are with the Institute of Advanced Sciences, Yokohama National University, 79-5 Tokiwadai, Hodogaya, Yokohama 240-8501, Japan (e-mail: takeuchi-naoki-kx@ynu.ac.jp; suzuh@ynu.ac.jp).

T. Yamae is with the Department of Electrical and Computer Engineering, Yokohama National University, 79-5 Tokiwadai, Hodogaya, Yokohama 240-8501, Japan (e-mail: yamae-taiki-yw@ynu.jp).

N. Yoshikawa is with the Institute of Advanced Sciences, Yokohama National University, 79-5 Tokiwadai, Hodogaya, Yokohama 240-8501, Japan; and also with the Department of Electrical and Computer Engineering, Yokohama National University, 79-5 Tokiwadai, Hodogaya, Yokohama 240-8501, Japan (e-mail: nyoshi@ynu.ac.jp).

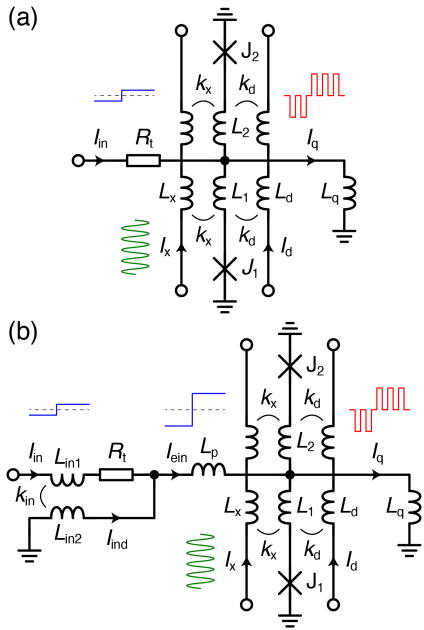

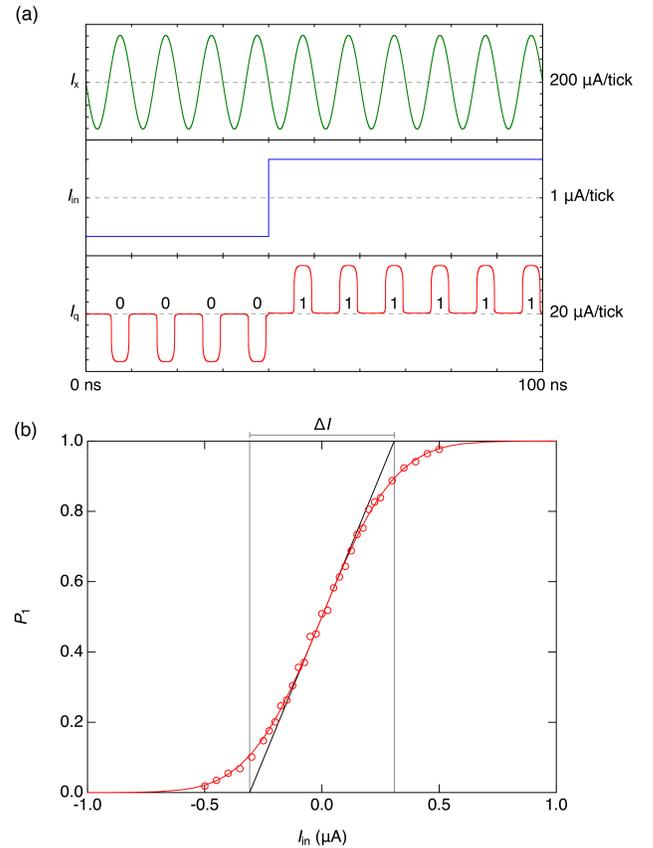

Fig. 1. Adiabatic quantum-flux-parametron (AQFP) comparators. (a) Conventional design. (b) Proposed design using a current transformer (CT). The critical current of $J_1$ and $J_2$ is 50 µA. $L_1 = L_2 = 1.59$ pH, $L_q = 7.92$ pH, and $R_t = 50$ Ω. $L_{in1} = 2.63$ nH, $L_{in2} = 30.0$ pH, $L_p = 1.77$ pH, and $k_{in} = 0.662$.

that $L_1 = L_2$ and the critical current of $J_1$ is the same as that of $J_2$. While $I_x$ is increasing, $J_1$ ($J_2$) switches for a positive (negative) $I_{in}$, and then a positive (negative) output current $I_q$ appears through a load inductor $L_q$, which represents a logical value of 1 (0). Consequently, the comparator digitizes $I_{in}$ with a reference current of 0 µA and a sampling frequency ($f_s$) equal to the frequency of $I_x$. Note that the reference current can be changed by applying an offset current to the input port [9].

Figure 2(a) shows the simulation results for typical waveforms with regard to the comparator for $f_s = 100$ MHz, where the gray dashed lines represent a zero for each waveform. In the present study, numerical simulation is conducted using a Josephson circuit simulator, JSIM [24], with the circuit parameters shown in the caption of Fig. 1 and junction parameters for the AIST 10 kA/cm² Nb high-speed standard process (HSTP) [13]. In Fig. 2(a), $I_{in}$ is correctly sampled and digitized in synchronization with $I_x$. Without thermal fluctuation, the comparator can digitize an arbitrarily small $I_{in}$, but, in practice, the sensitivity of the comparator is limited by thermal noise [17]. Figure 2(b) shows the simulation results for the probability of switching to a logical value of 1 ($P_1$) as a function of $I_{in}$ for $f_s = 100$ MHz and $T = 4.2$ K, where $T$ is the operating temperature. In this simulation, thermal fluctuation is taken into account by adding thermal noise current sources to $J_1$, $J_2$, and $R_t$ in parallel [25, 26]. In Fig. 2(b), the markers show the simulation results, and the line represents a fitting curve. Since the thermal noise currents follow a Gaussian distribution, the following fitting curve was used [27]:

$$P_1 = \frac{1}{2} + \frac{1}{2}\mathrm{erf}\left(\sqrt{\pi}\frac{I_{in}}{\Delta I}\right), \quad (1)$$

where $\Delta I$ is called the gray zone width and represents the sensitivity. Figure 2(b) shows that whereas the comparator can digitize $I_{in}$ correctly with a high probability for large $|I_{in}|$, the uncertainty in the sampling process increases as $|I_{in}|$ approaches 0 µA (since the comparator absorbs heat and increases entropy [28]).

Fig. 2. Transient analysis results for $f_s = 100$ MHz. (a) Typical waveforms. (b) Probability of switching to a logical value of 1 as a function of the input current at 4.2 K. $\Delta I$ represents the sensitivity of the comparator.

B. High-sensitivity Comparator

In order to reduce $\Delta I$ and improve the sensitivity, we propose the high-sensitivity comparator shown in Fig. 1(b), where a CT composed of $L_{in1}$, $L_{in2}$, and $k_{in}$ is inserted at the input port. Assuming that a parasitic inductance $L_p$ and the equivalent inductance of the AQFP part are much smaller than $L_{in2}$, an induced current $I_{ind} \approx k_{in}(L_{in1}L_{in2})^{0.5}I_{in}/L_{in2} = k_{in}(L_{in1}/L_{in2})^{0.5}I_{in}$ is generated through $L_{in2}$, where $k_{in}(L_{in1}/L_{in2})^{0.5}$ denotes impedance conversion. As a result, an equivalent input current $I_{ein} = I_{in} + I_{ind}$ appears via $L_p$. Thus, the current ratio $n$ between $I_{ein}$ and $I_{in}$ is given by:

$$n = \frac{I_{ein}}{I_{in}} \approx 1 + k_{in}\sqrt{\frac{L_{in1}}{L_{in2}}}. \quad (2)$$

This equation indicates that $n$ increases by increasing the inductance ratio $L_{in1}/L_{in2}$, which equivalently improves the sensitivity of the comparator. Based on the physical layout design shown later [see Fig. 4(b)], the parameters regarding the CT are as follows: $L_{in1} = 2.63$ nH, $L_{in2} = 30.0$ pH, and $k_{in} = 0.662$, which were extracted using an inductance extractor, InductEx [29]. This parameter set achieves an $n$ of approximately 7 using

Eq. (2), which was actually found to be 6.3 by numerical simulation.

In order to verify the benefit of the CT, we compare $\Delta I$ for the conventional and proposed comparators through numerical simulation. Figure 3(a) shows the simulation results for $\Delta I$ as a function of $f_s$ for $T = 4.2$ K, where the conventional comparator uses two damping conditions for comparison. The green triangle markers represent $\Delta I$ for the conventional comparator without a CT for a McCumber parameter [30] $\beta_c$ of 1 (i.e., $J_1$ and $J_2$ are critically damped by shunt resistors). The red square markers represent $\Delta I$ for the conventional comparator for a $\beta_c$ of approximately 180 (i.e., $J_1$ and $J_2$ are intrinsically damped by subgap resistance, without shunt resistors). The blue round markers represent $\Delta I$ for the proposed comparator with a CT for a $\beta_c$ of approximately 180. The three lines represent the fitting curves for the simulated $\Delta I$. At this moment, it is not clear how $\Delta I$ for an AQFP comparator depends on $f_s$ because the previous studies on single-flux-quantum (SFQ) and quantum-flux-parametron (QFP) comparators [31–37] are not applicable to AQFP comparators; whereas the previous comparators assume non-adiabatic switching from a meta-stable state to a stable state, AQFP comparators perform adiabatic switching (in which the comparator is always in the stable state during a switching process). Thus, we used a simple fitting function of $af_s^b + c$, where $a$, $b$, and $c$ are the fitting parameters. Here, $(a, b, c)$ is (2.00, 0.289, 0.202), (0.919, 0.315, 0.186), and (0.129, 0.329, 0.0382) for the conventional comparator with $\beta_c = 1$, that with a $\beta_c$ of approximately 180, and the proposed comparator with a $\beta_c$ of approximately 180, respectively. Figure 3(a) shows that the sensitivity of an AQFP comparator improves by increasing $\beta_c$ and/or adopting a CT, and that the proposed comparator achieves a sub-100 nA sensitivity for $f_s$ lower than approximately 100 MHz. This $\Delta I$ is much smaller than that of typical SFQ comparators (a few µA at 4.2 K for low frequencies) [37]. The improvement of $\Delta I$ by the CT for $f_s = 100$ MHz is 0.617 µA/0.0955 µA = 6.5, which is close to the simulated current ratio ($n = 6.3$). Higher $\beta_c$ achieves smaller $\Delta I$ because, as $\beta_c$ increases (i.e., the characteristic time of the Josephson junctions decreases), the state of the comparator approaches the potential minimum and becomes less sensitive to thermal fluctuation during a switching process.

Furthermore, we simulate the temperature dependence of $\Delta I$ for the proposed comparator because some detectors are operated at sub-Kelvin temperatures [20–23]. Figure 3(b) shows the simulation results for $\Delta I$ as a function of $T$ for $f_s = 100$ MHz. The markers show the simulation results, and the line is the fitting curve given by $\Delta I = dT^{0.5}$, where $d = 0.0462$ is the fitting parameter. This tendency (i.e., $\Delta I$ is proportional to $T^{0.5}$) agrees well with the temperature dependence of SFQ comparators [35, 36]. Here, $\Delta I$ for the proposed comparator reaches 15 nA at 100 mK and 4.6 nA at 10 mK. Note that even smaller $\Delta I$ is available by increasing $n$, as long as the time constant $L_{in1}/R_t$ is acceptable in light of the time scale for $I_{in}$. $L_{in1}/R_t$ for the current design is 52.6 ps, which is still much shorter than the typical rise time of $I_{in}$ for SSPDs (~ns).

## III. EXPERIMENTS

Lastly, we experimentally demonstrate the high sensitivity of the proposed comparator. Figure 4(a) shows the experimental setup for measuring $\Delta I$. An AQFP chip under test is placed inside a dipping probe and is immersed in liquid He (at 4.2 K). An arbitrary waveform generator (AWG: Tektronix, AFG3152C) applies a pair of excitation currents $I_{x1}$ and $I_{x2}$ with a phase separation of 90° to power and clock the AQFP comparators and peripheral circuits on the chip. The combination of $I_{x1}$, $I_{x2}$, and $I_d$ enables four-phase clocking [13]. Here, $I_{in}$ is applied via an on-chip 10 kΩ resistor to the comparator in order to reduce the noise between the room-temperature stage and cryogenic chip. The output voltage $V_{out}$ is amplified by a low-noise amplifier (LNA) and is then digitized by a pulse counter (PC: Stanford Research Systems, SR400), which counts the number of logical 1 values per second from the AQFP chip to calculate $P_1$. Figure 4(b) shows a micrograph of the AQFP comparators and peripheral circuits on the chip, which was fabricated using the HSTP. In order to clarify the benefit of a CT, both the conventional comparator (left-hand comparator in the micrograph) and the proposed comparator (right-hand comparator in the micrograph) are included. The inset in Fig. 4(b) is a close-up of the CT, in which $L_{in1}$ is formed by a seven-turn spiral inductor to increase the inductance ratio $L_{in1}/L_{in2}$. In order to isolate the comparators from the leakage bias currents from the readout dc SQUIDs [8], which convert signal currents in AQFP gates into $V_{out}$, buffer chains are inserted between the comparators and dc SQUIDs. In addition, an isolation inductor $L_{iso}$ (= 32.3 pH) is inserted between the comparator and buffer chain to mitigate the back action from the buffer chain to the comparator.

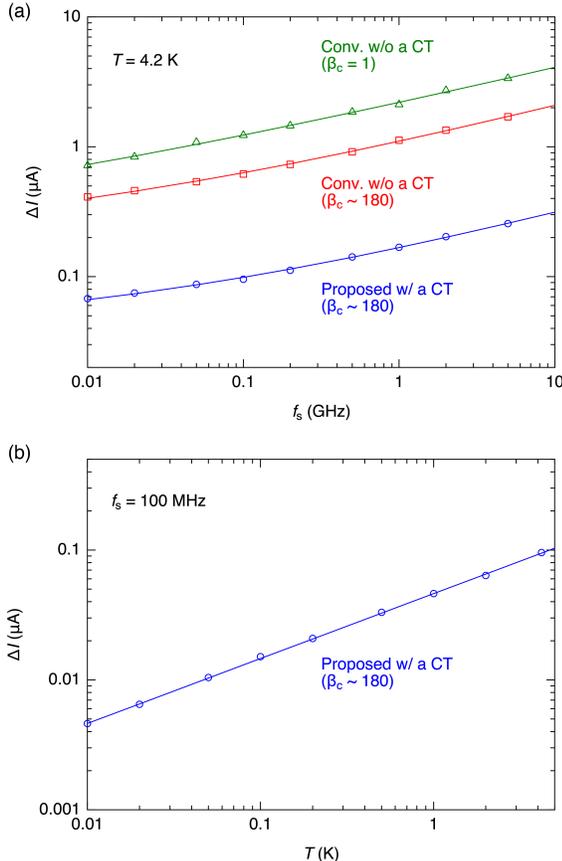

Fig. 3. Simulation of the sensitivity of AQFP comparators. (a) Gray zone width as a function of the sampling frequency. The use of a CT improves the sensitivity significantly. (b) Gray zone width as a function of the temperature. $\Delta I$ changes in proportion to $T^{0.5}$.

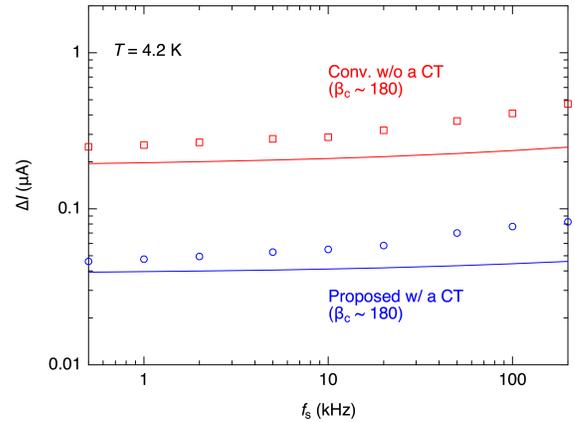

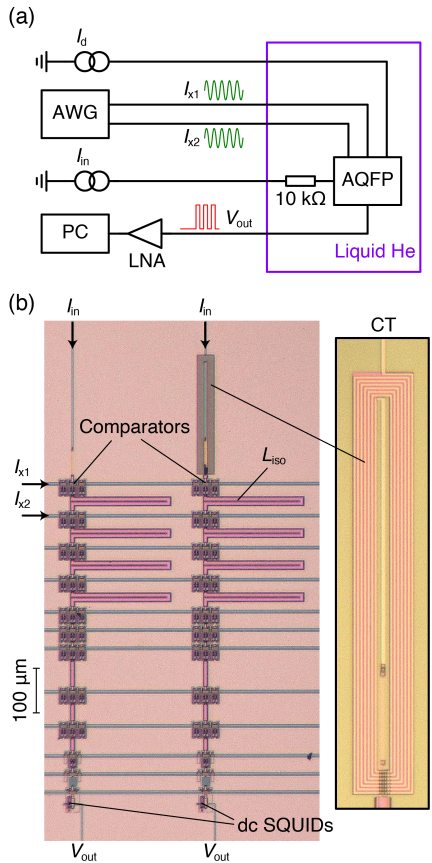

Fig. 4. (a) Experimental setup. The pulse counter (PC) counts the number of logical 1 values from the AQFP chip to calculate $P_1$. (b) Micrograph of the fabricated comparators.

Figure 5 shows the measurement results for $\Delta I$ as a function of $f_s$ for both the conventional and proposed comparators, where a low $f_s$ was used due to the narrow bandwidth of the experimental setup. In Fig. 5, the red square markers represent $\Delta I$ for the conventional comparator, and the blue round markers represent $\Delta I$ for the proposed comparator. The lines are the fitting curves given by the numerical simulation shown in Fig. 3(a). Figure 5 demonstrates that $\Delta I$ decreases significantly by adopting a CT and that the measurement results agree well with the simulation results. Note that a $\Delta I$ of as small as 46 nA was achieved by the proposed comparator at 500 Hz in the measurement. A slight difference between the measurement and simulation results would be due to the difference in noise condition. $I_{in}$ includes some noise in the experiment, whereas $I_{in}$ does not include any noise in the simulation.

## IV. CONCLUSION

In the present study, we proposed a high-sensitivity comparator using a CT, which increases the equivalent input current and reduces $\Delta I$. Numerical simulation showed that the proposed comparator achieves a $\Delta I$ of approximately six times smaller than that achieved by the conventional comparator. It was also found that $\Delta I$ decreases in proportion to $T^{0.5}$. Furthermore, in the experiment, we demonstrated that the proposed comparator achieves a $\Delta I$ of 46 nA at 4.2 K for $f_s$ = 500 Hz. The above results indicate the possibility of extending the application of AQFP logic to various cryogenic detectors. For instance, AQFP comparators may be able to be used as a readout interface for optical TESs [23], which generate signal currents of ~100 nA at sub-Kelvin temperatures. In future work, we will develop AQFP-based analog-to-digital convertors using high-sensitivity comparators for the readout of optical TESs.

Fig. 5. Measurement results for the sensitivity of AQFP comparators as a function of the sampling frequency at 4.2 K. The markers show the measurement results, and the lines are based on the simulation results. A $\Delta I$ of as small as 46 nA is achieved at 500 Hz by the proposed comparator.

## ACKNOWLEDGMENT

The present study was supported by the VLSI Design and Education Center (VDEC), University of Tokyo, in collaboration with Cadence Design Systems, Inc. The circuits were fabricated in the Clean Room for Analog-digital superconductiVITY (CRAVITY) of the National Institute of Advanced Industrial Science and Technology (AIST). We would like to thank C. J. Fourie for providing a 3D inductance extractor, InductEx, and H. Terai, S. Miki, F. China, D. Fukuda, and K. Hattori for useful discussions.